# Wavelet Time Shift Properties Integration with Support Vector Machines

Jaime Gómez[1], Ignacio Melgar[2] and Juan Seijas[3].

[1 2 3] Sener Ingeniería y Sistemas, S.A., Tres Cantos, Madrid. Spain.
[1] Escuela Politécnica Superior, Universidad Autónoma de Madrid.
[2 3] Departamento de Señales, Sistemas y Radiocomunicaciones, Universidad Politécnica de Madrid.
jaime.gomez@ii.uam.es
ignacio.melgar@sener.es
seijas@gc.ssr.upm.es

**Abstract.** This paper presents a short evaluation about the integration of information derived from wavelet non-linear-time-invariant (non-LTI) projection properties using Support Vector Machines (SVM). These properties may give additional information for a classifier trying to detect known patterns hidden by noise. In the experiments we present a simple electromagnetic pulsed signal recognition scheme, where some improve is achieved with respect to previous work. SVMs are used as a tool for information integration, exploiting some unique properties not found in neural networks.

## 1. Introduction

In previous work we have introduced a new algorithm to detect the presence (or absence) of electromagnetic signals using optimum theoretic discriminators [1] and Support Vector Machines [2] applied to wavelet transform output. This approach performs 15 dB better than previous algorithms using wavelets. The main advantage of our algorithm is its ability to integrate huge amounts of unrelated information, i.e., information coming from different sources (see figure 1).

However, this algorithm needed a time search process to be sure that in the case a signal is emitted, our system will process all its energy in at least one window. In [3] we introduced a valid time search algorithm and gave some hints about its ability to process greater amounts of information at low computational costs.

In this paper we focus on taking advantage of the time search process itself to improve the probability of detection ($P_d$) and probability of false alarm ($P_{fa}$). The wavelet transformation shows how time variant properties can be exploited using Machine Learning (ML) tools. Support Vector Machines is a ML tool with remarkable properties [4], especially useful in this case because of its ability to easily impose greater error penalty on one of the classes only.

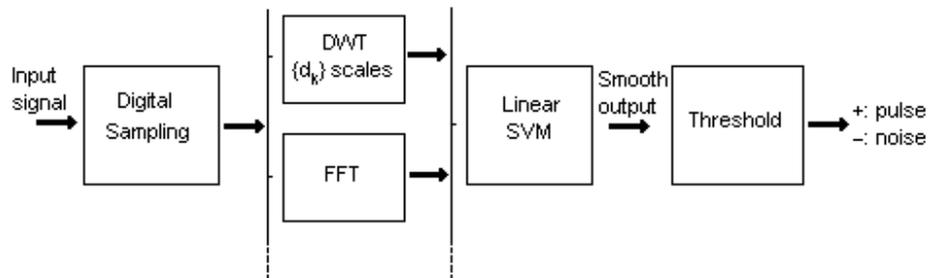

**Fig. 1.** Information processing scheme. Given a digitized signal, multiple transformations can be executed. Then a SVM will gather those features and classify them either as a pulsed signal or noise alone. Using a SVM, the threshold is defined as 0 (the threshold could be also identified as parameter *b*)

## 2. Wavelet properties of interest

Almost all existing signals can be expressed by a wavelet transform. Wavelets are generated by the scaling and translation of a single prototype function called wavelet mother [5]. The result of this transform is a set of vectors (called scales) whose coefficients describe the behaviour of the input data with respect to time and frequency. The discrete wavelet transform offers high time resolution for low scales (high frequencies) and high frequency resolution for high scales (low frequencies).

One of the main attributes of wavelets transforms is that they are non-LTI, i.e., they are time variant. Given a system where for an input $x(n)$ the output is $y(n)$, it is said to be time invariant if, for a shifted input $x(n-n_0)$, the output of the system is $y(n-n_0)$, independently of the chosen time shift $n_0$.

Suppose we define a known pulsed signal being N samples long (the sampling rate is usually defined by the digitizing hardware resources available). Let $d_k$ be the wavelet scale used to analyse the data, D an integer number, $D < N$, and H the size of the processing window, $H \geq N$ (the unit is always one sample). Suppose we apply the simple wavelet transform to a window having a pulsed signal centered on it, and noise on both sides. Now let's shift the data in the window so as to have the pulsed signal D samples away from the center, and let's calculate the wavelet transformation again (see figure 2). Of course, both windows output coefficients will be different, but as wavelets are non-LTI, those sets of coefficients will not have a direct relationship between them. Furthermore, even though they share the same source (the pulsed signal placed somewhere inside the window), they do not bear the same information.

The main difference between processed windows is that they hold different projections about one same reality. Each projection is not complete, it losses information. On the other hand, because of time variance, nearby windows have some degree of complementary information, which can be used to upgrade the overall picture about the input data.

For instance, in table 1 a cross-correlation coefficients matrix can be seen, relating different projections of the same pulse using different shifts (values of D), as will be defined on the experiments section.

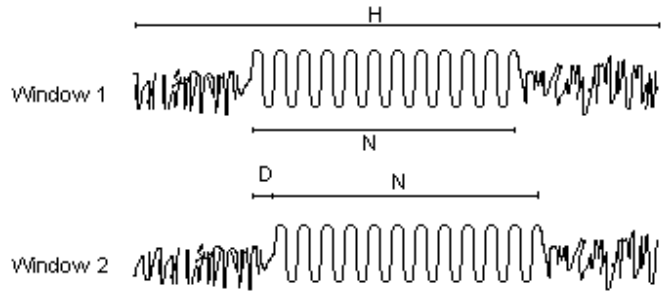

**Fig. 2.** Input data (somehow exaggerated for visibility purposes). For both input data the x-axis is time and y-axis is received power. The size of the computed window is H, the size of the pulsed signal is N and the difference between both snapshots is D. Both raw signals have the same amount of basic class information, but after wavelet processing projections will be complementary

**Table 1.** Cross-correlation coefficients matrix for 10000 pulse observations and three variables (0-shift, 11-shift and 23-shift).

$$\begin{pmatrix} 1.0000 & 0.9123 & 0.9573 \\ 0.9123 & 1.0000 & 0.8977 \\ 0.9573 & 0.8977 & 1.0000 \end{pmatrix}$$

Note that although the three variables are clearly not independent, they are not completely correlated, that is, all three variables hold some small portion of individual, unshared information. Therefore, as we are able to process this kind of data together efficiently using ML algorithms, we can be sure to obtain better results than any of the three sources alone. Moreover, the more incomplete pulse information sources we use, the better results we will obtain. As more variables are introduced in the model, new uncorrelated information will be harder to find, but nevertheless there will always be some improvement.

## 3. Integration Algorithm

### 3.1 Time-Shift Scheme

The time search algorithm described in [3] gives some constraints to the parameters N, D and H introduced in the previous section. As the wavelet calculation is the most expensive step throughout the algorithm, it is wise to reduce the size of the window H to the minimum. Therefore we set H=N (preferable $N=2^n$). To guarantee we will always be able to obtain the complete N-samples pulse in at least one N-samples

window we can skip no sample, so the shift distance between two consecutive
windows should be set to 1. Nevertheless, depending on the scale used and the pulse
form itself, the shift distance between consecutive windows could be lightly increased
so as to minimize computational cost.

Note that, unlike the general algorithm, these parameter constraints define one
complete-pulse and many incomplete-pulse projections. Even though the incomplete-
pulse windows should bear less information as D gets higher, they contribute with
some degree of classification upgrade, as will be observed in the experiments section.

After defining these concepts, we have to choose which shifts distances with respect
to the basic, centered-pulse window (D values) will be used for information
integration in the last phase of the algorithm. Such choice depends on several factors:
pulse frequency, pulse size, sampling rate, and wavelet scales used in the first steps.
Usually consecutive windows (one sample shift away) projections would be very
much alike. Two similar projections are of no use together, they bear no more
information as a whole than separately. In our projection choice, we need to have a
balance between non-similarity (not too close) and usefulness (having a big chunk of
source data). The smallest the shift with respect to the complete-pulse window the
better, but not so close as to have redundant projections. Also, as wavelets are built
using powers of two, it seems not wise to use even shifts.

Having these set of experimental rules in mind, in our experiments we chose as the
shift D prime numbers around one percent away from the complete-pulse window
onwards.

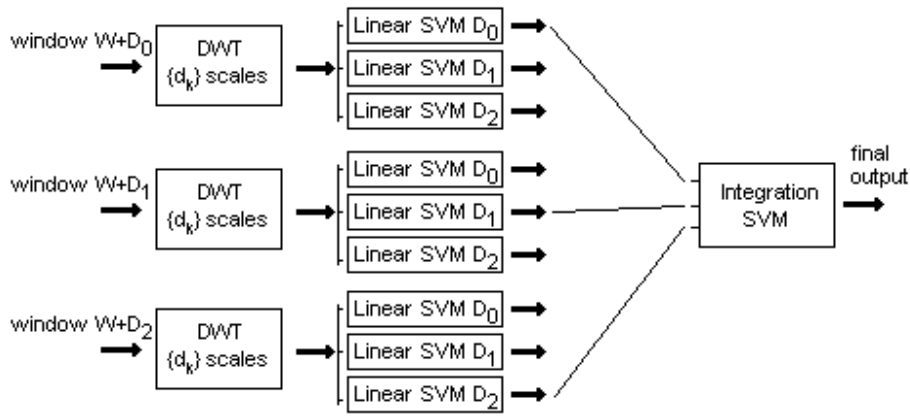

**Fig. 3.** Integration procedure for information coming from different projections.

### 3.2 SVM Integration

To integrate the information coming from different sources (see figure 3) we need a
tool able to perform a statistical evaluation on how a separator decision function can be sustained on more than one variable or feature. SVM are a very nice tool at founding the best separator surface for such a scenario. Using non-linear kernels we are able to generate non-linear relationships between input features that may adapt better to the statistics behind the sources of information.

Given two variables, if they are completely correlated, then they share all the information, i.e., one of them is useless. On the other hand, if two variables defined as two different representations of one single event are not completely correlated, then it is highly probable that a ML process will be able to extract different unshared information from both of them, improving classification rates.

Using the cross-correlation coefficients matrix of the input data we can perform some rough estimates about how uncorrelated are each variable against the other sources of information. In table 2 another matrix is shown similarly to table 1. In this table we have added two more variables, defined as 37-shift and 53-shift. Note that both new variables are highly correlated (its relation coefficient is 0.9951, very close to 1), and one of them would be useless in the classification function. Nevertheless, they can still have some sort of information not present in the other three representations of the same event. They are useful, but only one of them.

**Table 2.** Cross-correlation coefficients matrix of 10000 pulse observations and five variables (0-shift, 11-shift, 23-shift, 37-shift and 53-shift).

$$\begin{pmatrix} 1.0000 & 0.9123 & 0.9573 & 0.9209 & 0.9201 \\ 0.9123 & 1.0000 & 0.8977 & 0.9131 & 0.9171 \\ 0.9573 & 0.8977 & 1.0000 & 0.9474 & 0.9449 \\ 0.9209 & 0.9131 & 0.9474 & 1.0000 & 0.9951 \\ 0.9201 & 0.9171 & 0.9449 & 0.9951 & 1.0000 \end{pmatrix}$$

To integrate all this information, we used a non-linear inhomogeneous polynomial kernel of degree two. Greater degree polynomial kernels (up to the number of input features) performed worse, so did the linear kernel to a lesser extent. Those non-linear features gave a flexibility to the decision function good enough to attain an optimum, but the main advantage of using the SVM was its capability to give different penalty cost to errors of each class. We used the two-fold C parameter (C+ and C–), which gave us the possibility to fulfil the very restricted requirements of $P_{fa}$ analysis. Neural Networks do not posses this capability incorporated to the basic training algorithm, and therefore it could not be used as such tool.

**3.3 Algorithm complexity**

Let us set the multiply-add as the basic operation. Let us define H as the initial vector size, W as the wavelet filter size and K as the wavelet scale, then the complexity of wavelet calculation is O(KWH). Let us also define S as the size of the coefficient vector at scale K and M as the number of linear SVM classifiers trained for different complete / incomplete pulse schemes, then linear classifiers execution is O(MS). The non-linear integration SVM uses very few input features (M), and therefore it can be easily described with a small set of support vectors (not much greater than M), using the reduced set approach by Burges [6].

Adding up these intermediate process, computational requirements are O(KWH) + O(MS) + O($M^2$). Therefore, the algorithm complexity remains bounded to the wavelet transform computational cost O(KWH). For a small additional cost we can obtain much better results, as can be observed throughout the experiments.

## 4. Experiments

Our experiments had the following setup: chirp pulse (see [1]), 1024 samples size (N); mother wavelet Daubechies 5, using d4 wavelet scale coefficients; white Gaussian noise with zero mean and deviation equals one; five linear detectors (SVMs) were trained such as to detect the complete pulse (named 0-shift), 11 noise samples plus 1013 pulse samples (named 11-shift), and a similar approach for 23 samples (named 23-shift), 37 samples (named 37-shift), 53 samples (named 53-shift) and having $P_{fa} = 10^{-3}$, as established in [2].

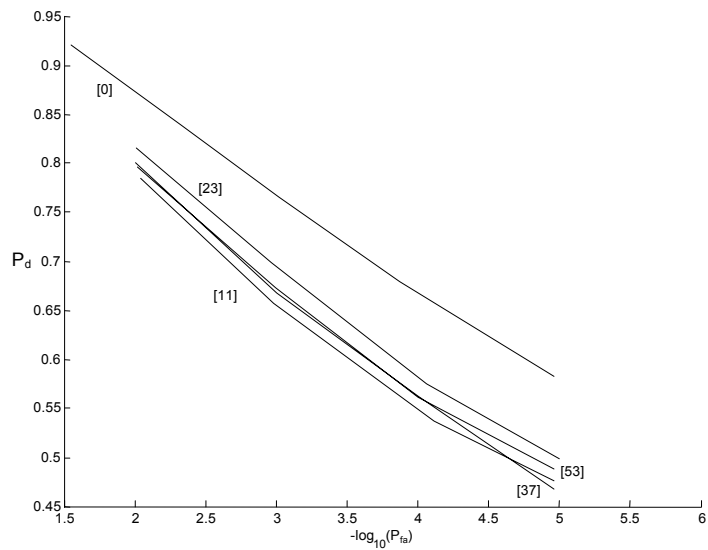

**Fig. 4.** Performance functions for individual wavelet+LinearSVM schemes for complete- and incomplete- pulse windows. The X axis corresponds to $-\log_{10}(P_{fa})$. The Y axis corresponds to the mean for $P_d$ results for SNR values between 0 and −15 dB.

Thus, for each window observation, we computed 5 similar Wavelet + Decision-function schemes. We generated two integration SVM, one having as inputs the 0-,

11- and 23-shift smooth outputs, and the other having 0-, 11-, 23-, 37-, 53-shift smooth outputs, all of them extracted from the corresponding windows. Figures are expressed as the probability of detection mean on some SNR interval with respect to desired probability of false alarm.

In figure 4 some interesting features observed on tables 1 and 2 are confirmed. The best performing scheme is the 0-shift (complete pulse), but the second best performing is not the 11-shift one, which is the next having the bigger piece of signal. This projection bears less information individually, although it may still have unshared features useful for a multiple scheme SVM integration. Note that the size of the pulse varies slightly from one scheme to another. The smallest pulse size (the 53-shift) has around 95% of the pulse present. See also that both 37-shift and 53-shift projections have very similar performance.

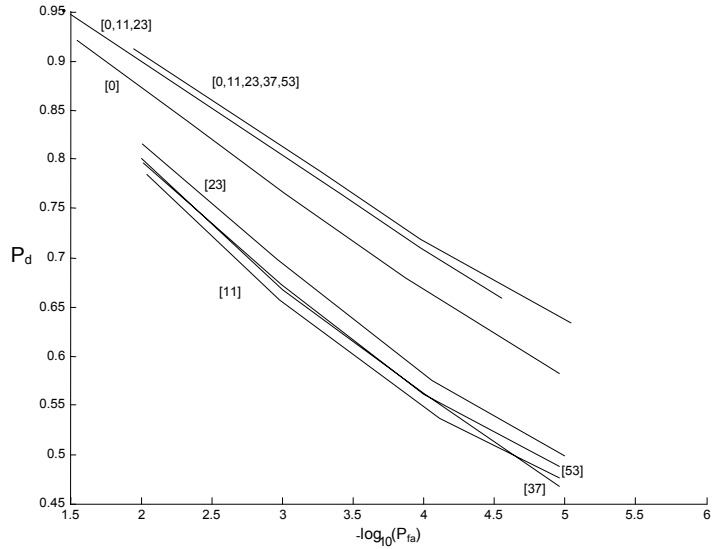

**Fig. 5.** Performance functions for multiple wavelet+LinearSVM plus integration SVM schemes for two sets of individual projections: [0,11,23] and [0,11,23,37,53]. Also the individual schemes of figure 4 are shown for comparison. The X axis corresponds to $-\log_{10}(P_{fa})$. The Y axis corresponds to the mean for $P_d$ results for SNR values between 0 and −15 dB.

Finally, in figure 5 we can see the effects of multiple projection integration. The $P_{fa}$ upgrade is a bit less than one order of magnitude, increasing as $P_{fa}$ requirements become harder. Note also how two added variables (37-shift and 53-shift), which were observed as having a great correlation between them and considerable less class information, are able to improve substantially our results.

## 5. Conclusions

In this paper we have seen a new unlimited line of information processing: the use of additional Wavelet + Decision-function scheme applied to previously determined incomplete pulsed signals. This new set of features provides the final model with uncorrelated information, upgrading the classification rates.

Non-LTI properties in wavelet transforms define different useful projections of reality, which yield unshared information. In our experiments we have analysed the case when multiple incomplete-pulse schemes are executed, but this algorithm can be used also on multiple complete-pulse schemes (need only H > N as defined on section 2), obtaining better overall results. In our experiments we wanted to emphasize the fact that even when processing less energy (shifted window), the resulting projection may have additional useful information to the complete pulse window projection.

Support Vector Machines is a great tool for information integration for linear as well as non-linear kernel functions. In this application we have confirmed the easy of use of SVMs as a Machine Learning tool regarding its capability to provide different weights for both classes, allowing the training system to comply with the otherwise difficult probability of false alarm rates.

Further analysis is needed to determine how an increased number of processing units in one multiple-source decision function will upgrade the system capabilities. We will also analyse in more depth how the shift number (we used small prime numbers only) affects the pulse projection.

## 6. Acknowledgements

This project is funded by Sener Ingeniería y Sistemas, in the frame of the Aerospace Division R&D program, reference P215903.## References

1. Melgar I., Gomez J., Seijas J.: Optimum Signal Linear Detector in the Discrete Wavelet Transform – Domain. World Scientific and Engineer Academy and Society Conference on Signal Processing, Computational Geometry, and Artificial Vision (ISCGAV'03), Rhodes, Greece, November 2003.
2. Gomez J., Melgar I., Seijas J., Andina D.: Sub-optimum Signal Linear Detector Using Wavelets and Support Vector Machines. World Scientific and Engineer Academy and Society Conference on Automation and Information (ICAI'03), Tenerife, Spain, December 2003.
3. Gomez J., Melgar I., Seijas J.: "Upgrading pulse detection with time shift properties using wavelets and Support Vector Machines". World Automation Congress (WAC'04), Seville, Spain, to appear.
4. Burges C.: A Tutorial on Support Vector Machines for Pattern Recognition. Knowledge Discovery and Data Mining, 2(2), pp 121-167, 1998.

5. Mallat, S.: A theory for multiresolution signal decomposition: the wavelet representation. IEEE Trans. Pattn Anal. Mach. Intell., 11, 674–693, 1989.
6. Burges C.: Simplified Support Vector Decision Rules. In L. Saitta, editor, *Proc. 13th International Conference on Machine Learning*, pages 71-77, San Mateo, CA, 1996. Morgan Kaufmann. B. Schölkopf .